\documentclass[twocolumn]{aastex631}
\usepackage{graphicx,amsmath,amsthm}

\shorttitle{Magnetic and Rotational Evolution of $\rho$~CrB}
\shortauthors{Metcalfe et al.}

\journalinfo{The Astrophysical Journal}

\begin{document}

\title{\bf Magnetic and Rotational Evolution of $\rho$~CrB from Asteroseismology with TESS}

\author[0000-0003-4034-0416]{Travis S.~Metcalfe}
\affiliation{White Dwarf Research Corporation, 9020 Brumm Trail, Golden, CO 80403, USA}
\affiliation{Space Science Institute, 4765 Walnut St., Suite B, Boulder, CO 80301, USA}

\author[0000-0002-4284-8638]{Jennifer L.~van~Saders}
\affiliation{Institute for Astronomy, University of Hawai`i, 2680 Woodlawn Drive, Honolulu, HI 96822, USA}

\author[0000-0002-6163-3472]{Sarbani Basu}
\affiliation{Department of Astronomy, Yale University, PO Box 208101, New Haven, CT 06520-8101, USA}

\author[0000-0002-1988-143X]{Derek Buzasi}
\affiliation{Department of Chemistry and Physics, Florida Gulf Coast University, 10501 FGCU Blvd S, Fort Myers, FL 33965}

\author[0000-0002-0210-2276]{Jeremy J.~Drake}
\affiliation{Harvard-Smithsonian Center for Astrophysics, Cambridge, MA 02138, USA}

\author[0000-0002-4996-0753]{Ricky Egeland}
\affiliation{High Altitude Observatory, National Center for Atmospheric Research, P.O. Box 3000, Boulder, CO 80307-3000, USA}

\author[0000-0001-8832-4488]{Daniel Huber}
\affiliation{Institute for Astronomy, University of Hawai`i, 2680 Woodlawn Drive, Honolulu, HI 96822, USA}

\author[0000-0001-7032-8480]{Steven H.~Saar}
\affiliation{Harvard-Smithsonian Center for Astrophysics, Cambridge, MA 02138, USA}

\author[0000-0002-3481-9052]{Keivan G.~Stassun} 
\affiliation{Vanderbilt University, Department of Physics \& Astronomy, 6301 Stevenson Center Lane, Nashville, TN 37235, USA}

\author[0000-0002-4773-1017]{Warrick H.~Ball}
\affiliation{School of Physics \& Astronomy, University of Birmingham, Edgbaston, Birmingham B15 2TT, UK}
\affiliation{Stellar Astrophysics Centre, Aarhus University, Ny Munkegade 120, DK-8000 Aarhus C, Denmark}

\author[0000-0002-4588-5389]{Tiago L.~Campante}
\affiliation{Instituto de Astrof\'{\i}sica e Ci\^{e}ncias do Espa\c{c}o, Universidade do Porto,  Rua das Estrelas, 4150-762 Porto, Portugal}
\affiliation{Departamento de F\'{\i}sica e Astronomia, Universidade do Porto, Rua do Campo Alegre, s/n, 4169-007 Porto, Portugal}

\author[0000-0002-3020-9409]{Adam J.~Finley}
\affiliation{Department of Astrophysics-AIM, University of Paris-Saclay and University of Paris, CEA, CNRS, Gif-sur-Yvette Cedex 91191, France}

\author[0000-0003-3061-4591]{Oleg Kochukhov}
\affiliation{Department of Physics and Astronomy, Uppsala University, Box 516, SE-75120 Uppsala, Sweden}

\author[0000-0002-0129-0316]{Savita Mathur}
\affiliation{Instituto de Astrof\'{\i}sica de Canarias, E-38205 La Laguna, Tenerife, Spain}
\affiliation{Dpto. de Astrof\'{\i}sica, Universidad de La Laguna, E-38206 La Laguna, Tenerife, Spain}

\author[0000-0002-1299-1994]{Timo Reinhold}
\affiliation{Max-Planck-Institut f\"ur Sonnensystemforschung, Justus-von-Liebig-Weg 3, 37077, G\"ottingen, Germany}

\author[0000-0001-5986-3423]{Victor~See}
\affiliation{University of Exeter, Department of Physics \& Astronomy, Stocker Road, Exeter, Devon, EX4 4QL, UK}

\author{Sallie Baliunas}
\affiliation{Harvard-Smithsonian Center for Astrophysics, Cambridge, MA 02138, USA}

\author{Willie Soon}
\affiliation{Harvard-Smithsonian Center for Astrophysics, Cambridge, MA 02138, USA}

\begin{abstract}
During the first half of main-sequence lifetimes, the evolution of rotation and magnetic activity in solar-type stars appears to be strongly coupled. Recent observations suggest that rotation rates evolve much more slowly beyond middle-age, while stellar activity continues to decline. We aim to characterize this mid-life transition by combining archival stellar activity data from the Mount Wilson Observatory with asteroseismology from the {\it Transiting Exoplanet Survey Satellite} (TESS). For two stars on opposite sides of the transition (88~Leo and $\rho$~CrB), we independently assess the mean activity levels and rotation periods previously reported in the literature. For the less active star ($\rho$~CrB), we detect solar-like oscillations from TESS photometry, and we obtain precise stellar properties from asteroseismic modeling. We derive updated X-ray luminosities for both stars to estimate their mass-loss rates, and we use previously published constraints on magnetic morphology to model the evolutionary change in magnetic braking torque. We then attempt to match the observations with rotational evolution models, assuming either standard spin-down or weakened magnetic braking. We conclude that the asteroseismic age of $\rho$~CrB is consistent with the expected evolution of its mean activity level, and that weakened braking models can more readily explain its relatively fast rotation rate. Future spectropolarimetric observations across a range of spectral types promise to further characterize the shift in magnetic morphology that apparently drives this mid-life transition in solar-type stars.
\end{abstract}


\NewPageAfterKeywords
\section{Introduction}\label{sec1} 

Young solar-type stars typically have strong magnetic fields with complex morphologies, like the closed loops surrounding active regions on the Sun \citep{Garraffo2018}. After about 50~Myr, the underlying stellar dynamo mechanism apparently becomes efficient at organizing the magnetic field on larger scales. The emergence of this large-scale organization has important consequences for the strong coupling between rotation and magnetic activity during the first half of stellar main-sequence lifetimes \citep{Skumanich1972}. The physical mechanism that produces this coupling is known as magnetic braking. Charged particles in a stellar wind are entrained in the magnetic field out to a critical distance known as the Alfv\'en radius, carrying away stellar angular momentum in the process. Most of the angular momentum that is lost from magnetic braking can be attributed to the largest scale components of the field, which have a longer effective lever-arm and more open field lines where the stellar wind can escape \citep{Reville2015, Garraffo2016, See2019}.

Middle-aged stars often have some of the clearest stellar activity cycles \citep{Brandenburg2017}. This may be a consequence of their slower rotation rates, which either fail to excite a second dynamo in the near surface shear layer \citep{BohmVitense2007}, or yield activity cycle periods that are much longer than the currently available data sets \citep{Baliunas1995}. Not long after rotation becomes slow enough to produce monoperiodic activity cycles ($P_{\rm rot}\sim 20$~days for solar analogs), it becomes too slow to imprint substantial Coriolis forces on the global convective patterns \citep{Featherstone2016}. This leads to a disruption of the solar-like pattern of differential rotation (i.e.\ faster at the equator and slower at the poles), and a gradual loss of shear to drive the organization of large-scale field by the global dynamo. The observational consequences of this mid-life transition include nearly uniform rotation in older stars \citep{Benomar2018}, weakened magnetic braking that temporarily stalls the rotational evolution \citep{vanSaders2016, Hall2021}, and a gradual decline in stellar activity until the cycles disappear entirely \citep{Metcalfe2016, Metcalfe2017}.

\cite{Metcalfe2019} recently tested this new understanding of magnetic stellar evolution using spectropolarimetric measurements of two stars with activity levels on opposite sides of the proposed mid-life transition. The more active star 88~Leo has a rotation period near 14 days and exhibits clear activity cycles, while the less active star $\rho$~CrB has a rotation period near 17 days and shows constant activity over several decades of monitoring \citep{Baliunas1995, Baliunas1996}. The snapshot observations with the Potsdam Echelle Polarimetric and Spectroscopic Instrument \citep[PEPSI,][]{Strassmeier2015} on the Large Binocular Telescope (LBT) appeared to confirm the predicted loss of large-scale magnetic field. The data produced a clear detection of a nonaxisymmetric dipole field in 88~Leo, and an upper limit on the dipole field strength in $\rho$~CrB that was well below what would be expected from its relative activity level---suggesting that most of the field was concentrated in smaller spatial scales. The age of $\rho$~CrB from gyrochronology (a lower limit on the actual age with weakened magnetic braking) was reported to be $2.5\pm0.4$~Gyr by \cite{Barnes2007}, while other age indicators suggested that it is substantially more evolved \citep{Valenti2005, Mamajek2008}.

In this paper, we aim to characterize the proposed magnetic transition by combining archival stellar activity data from the Mount Wilson Observatory (MWO) with asteroseismology from the {\it Transiting Exoplanet Survey Satellite} \citep[TESS,][]{Ricker2014}. In Section~\ref{sec2}, we reanalyze the complete MWO data sets for $\rho$~CrB and 88~Leo to assess their mean activity levels and rotation periods, we use TESS photometry to search for solar-like oscillations, we obtain X-ray luminosities to help constrain mass-loss rates, and we adopt additional constraints on the stellar properties using published spectroscopy, photometry, and astrometry. In Section~\ref{sec3}, we detect a signature of solar-like oscillations in $\rho$~CrB, and we derive precise stellar properties from asteroseismic modeling. In Section~\ref{sec4}, we assess the compatibility of the observations with an activity-age relation for solar analogs \citep{LorenzoOliveira2018}, and we estimate the magnetic braking torque using a simple wind modeling prescription. In Section~\ref{sec5}, we attempt to match the observations with rotational evolution models that assume either standard spin-down or weakened magnetic braking. Finally, we summarize and discuss our results in Section~\ref{sec6}, concluding that the asteroseismic age of $\rho$~CrB is consistent with the expected evolution of its mean activity level, and that weakened braking models can more readily explain its relatively fast rotation rate.

 \begin{figure*}[p]
 \centering\includegraphics[width=5.5in]{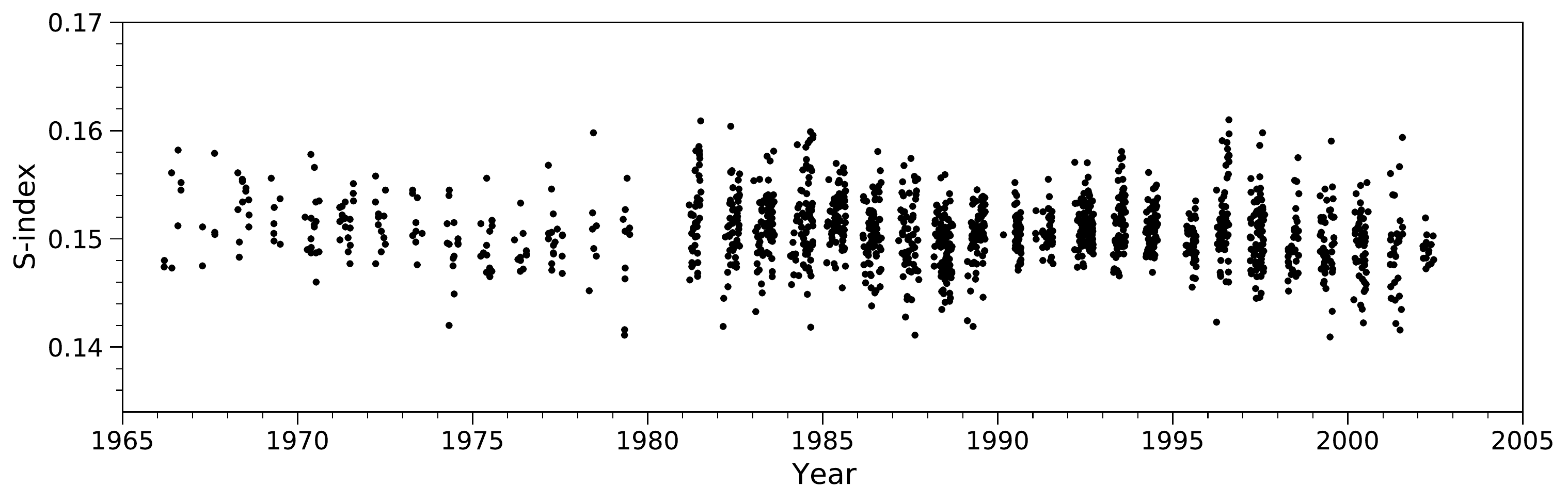} 
 \hspace*{2pt}\centering\includegraphics[width=5.35in]{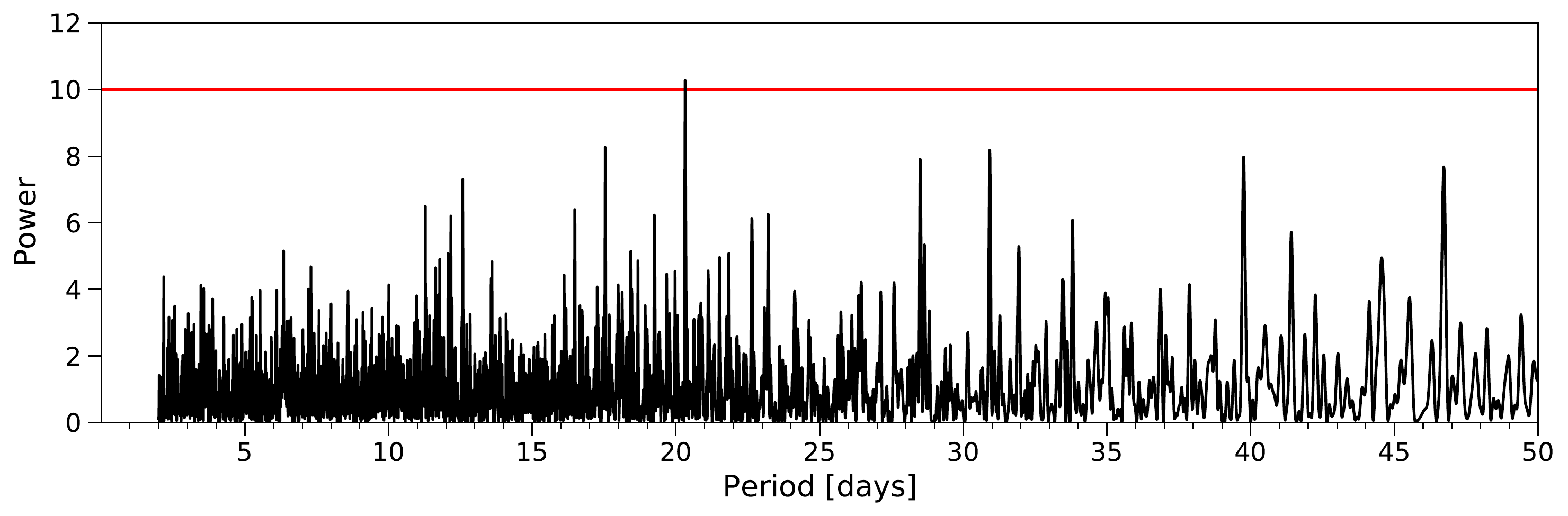}\vspace*{12pt}
 \includegraphics[width=5.5in]{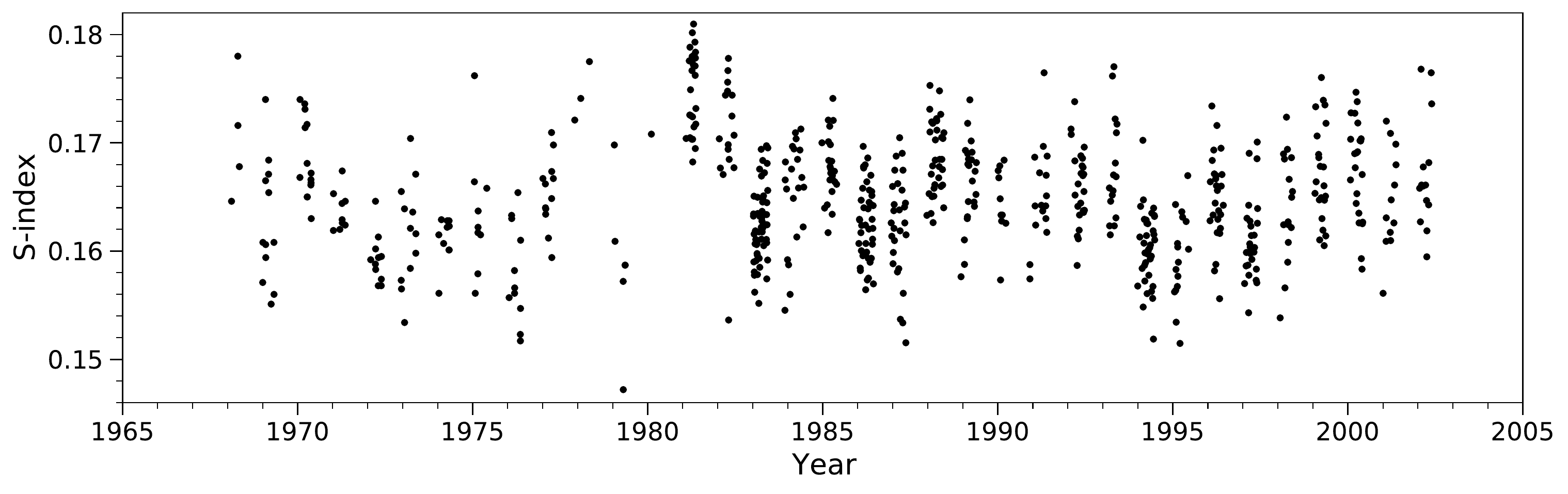} 
 \hspace*{2pt}\centering\includegraphics[width=5.35in]{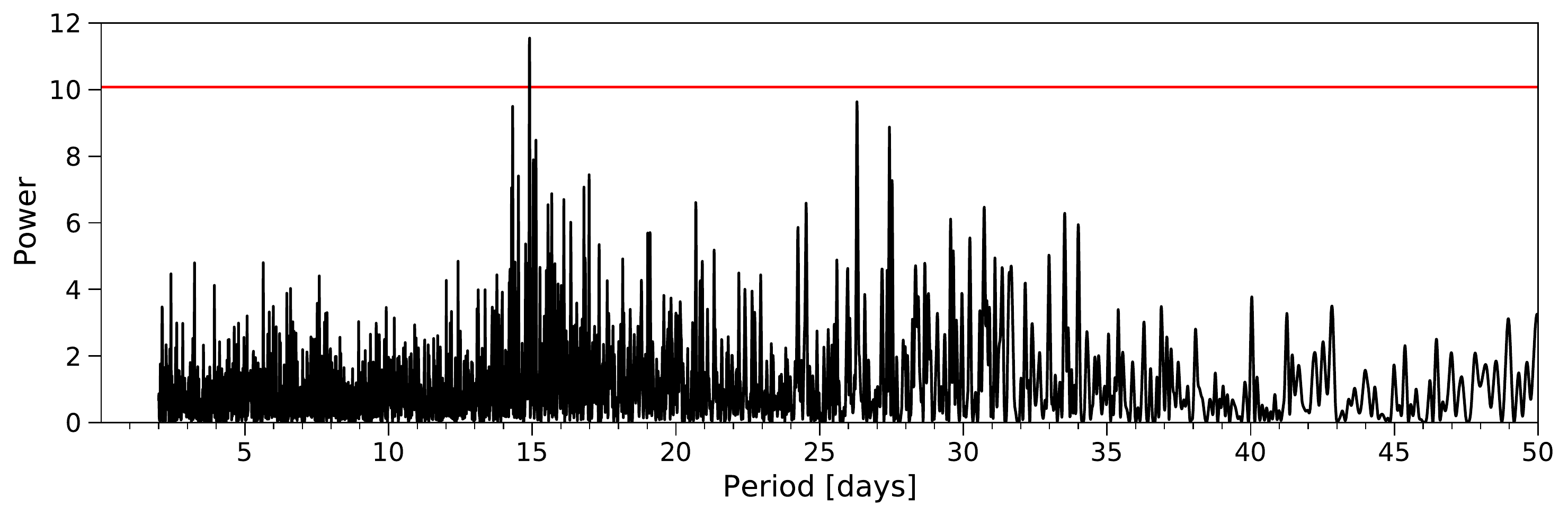} 
 \caption{Time series and Lomb-Scargle periodograms for $\rho$~CrB (top two panels) and 88~Leo (bottom two panels) showing the recovered rotation signals of 20.3 days and 15.0 days, respectively.  Red lines in the periodograms indicate the 5\% false alarm probability calculated from a Monte Carlo process. (The data used to create this figure are available).\label{fig1}}
 \end{figure*} 

\section{Observations}\label{sec2}

\subsection{Mount Wilson HK data}\label{sec2.1} 

Both $\rho$~CrB and 88~Leo have synoptic $S$-index time series from the MWO HK Project, ranging from near the beginning of the program in 1966 to its termination in 2003 (see Figure~\ref{fig1}). The MWO $S$-index measures the ratio of emission from 1~\AA{} cores of the Ca {\sc ii} H \& K lines to the sum of two nearby 20~\AA{} pseudo-continuum bandpasses \citep{Vaughan1978}. Such measurements are routinely used in studies of magnetic activity cycles and stellar rotation \citep[e.g.][]{Baliunas1996, Donahue1996}. Our analysis of the complete MWO time series gives a mean $S$-index of 0.1508 for $\rho$~CrB and 0.1655 for 88~Leo, in agreement with previous averages from the subset of data analyzed in \citet{Baliunas1995}. Adopting the spectroscopic temperatures from \cite{Brewer2016} and the activity scale from \cite{LorenzoOliveira2018}, we find $\log R'_{\rm HK}[T_{\rm eff}] = -5.177 \pm 0.015$ for $\rho$~CrB and $-4.958 \pm 0.015$ for 88~Leo (see Section~\ref{sec4}).

We applied the Lomb-Scargle periodogram to the entire time series as well as seasonal bins in order to search for rotational signals.  We took signals with a false alarm probability (FAP) less than 5\% to be statistically significant. The FAP is defined as the probability that a peak in the periodogram is due to Gaussian noise \citep{Horne1986}, and we have calculated the FAP using that definition explicitly in a Monte Carlo simulation of 100,000 trials.  In each trial, synthetic data of the same sampling cadence and standard deviation as the observational data are randomly drawn from the Gaussian distribution, and the Lomb-Scargle periodogram is computed.  The fraction of random trials generating periodogram peaks higher than the one obtained from the observational data is the FAP.  The uncertainty in the period is found by a similar Monte Carlo process where the observational data are moved within their 1\% uncertainty \citep{Baliunas1995} and the standard deviation of peak periods is computed.  Using this method, we find a rotation period of $20.3 \pm 1.8$ days for $\rho$~CrB (FAP\,=\,4.2\%) and $15.0 \pm 0.3$ days for 88~Leo (FAP\,=\,1.2\%) from the complete time series. Figure~\ref{fig1} shows the time series and Lomb-Scargle periodograms for both stars, with the 5\% FAP line computed from the Monte Carlo shown as a red line.  Single season analyses returned no significant peaks for $\rho$~CrB \citep[which is not unusual for ``flat activity'' stars,][]{Donahue1996}, and one season with a significant peak for 88~Leo, giving a rotation period of $14.3 \pm 0.8$ days (FAP\,=\,1.4\%) and confirming the global result.

Our rotation period for $\rho$~CrB is $\sim$2$\sigma$ longer than the 17~days found by \citet{Baliunas1996}, who used a subset of the MWO data and did not provide an uncertainty. However, our result agrees with \citet{Henry2000}  who used a longer subset of the MWO data ($\langle P_{\rm rot} \rangle = 19 \pm 2$~days, with seasonal values between 17--20~days), and with \citet{Fulton2016} who found 18.5~days from Keck observations. For 88~Leo, we find good agreement with the 14~day rotation period determined by \citet{Baliunas1996} and the 14.32~day period determined by \citet{Olah2009} from the complete MWO time series.

\subsection{TESS photometry}\label{sec2.2} 

TESS observed $\rho$~CrB in 2-minute cadence for a total of approximately 52 days during Sectors 24 and 25 of Cycle~2 (2020~Apr~15 – 2020~Jun~08). We downloaded the PDC-MAP SPOC light curve \citep{Jenkins2016},  but also derived our own light curve following the procedure described in \citet{Nielsen2020} and \citet{Buzasi2015} in hopes of improving on the noise level in the SPOC product. We treated sectors individually, masking cadences with nonzero quality flags. We then built a collection of single-pixel light curves for each pixel in the $25 \times 25$ pixel postage stamp. Our figure of merit for the quality of a light curve was the sum of the absolute values of the first-differenced light curve, generally a good proxy for high-frequency noise \citep{Nason2006}. Starting from the brightest pixel, we then added pixels one at a time to the light curve, choosing in each case the pixel that produced the largest decrease in our noise figure of merit, and continuing until light curve quality no longer improved. This process resulted in a somewhat larger aperture than that derived by the SPOC (114 pixels vs.\ 59 for Sector 24 and 108 pixels vs.\ 51 for Sector 25). The resulting light curves were then detrended of instrumental effects by fitting a second-order polynomial in $x$ and $y$ pixel location. We compared the resulting light curve to the SPOC product; improvement was modest but noticeable ($\sim$6\% decreased noise) at frequencies above 1~mHz, so we chose to use our light curve for the asteroseismic analysis in Section~\ref{sec3.1}.

We applied a similar photometric reduction algorithm to 88~Leo. TESS observed this star in 2-minute cadence for a total of approximately 27 days during Sector~22 of Cycle~2 (2020~Feb~18 – 2020~Mar~18). Once again, the process resulted in a somewhat larger aperture than that derived by the TESS SPOC (71 pixels vs.\ 36). After extraction and detrending, the noise level was lowered by approximately 15\% above 1~mHz.

\subsection{Spectral Energy Distribution}\label{sec2.3} 

\begin{figure}
 \centering\includegraphics[width=\columnwidth,trim=150 70 80 70,clip]{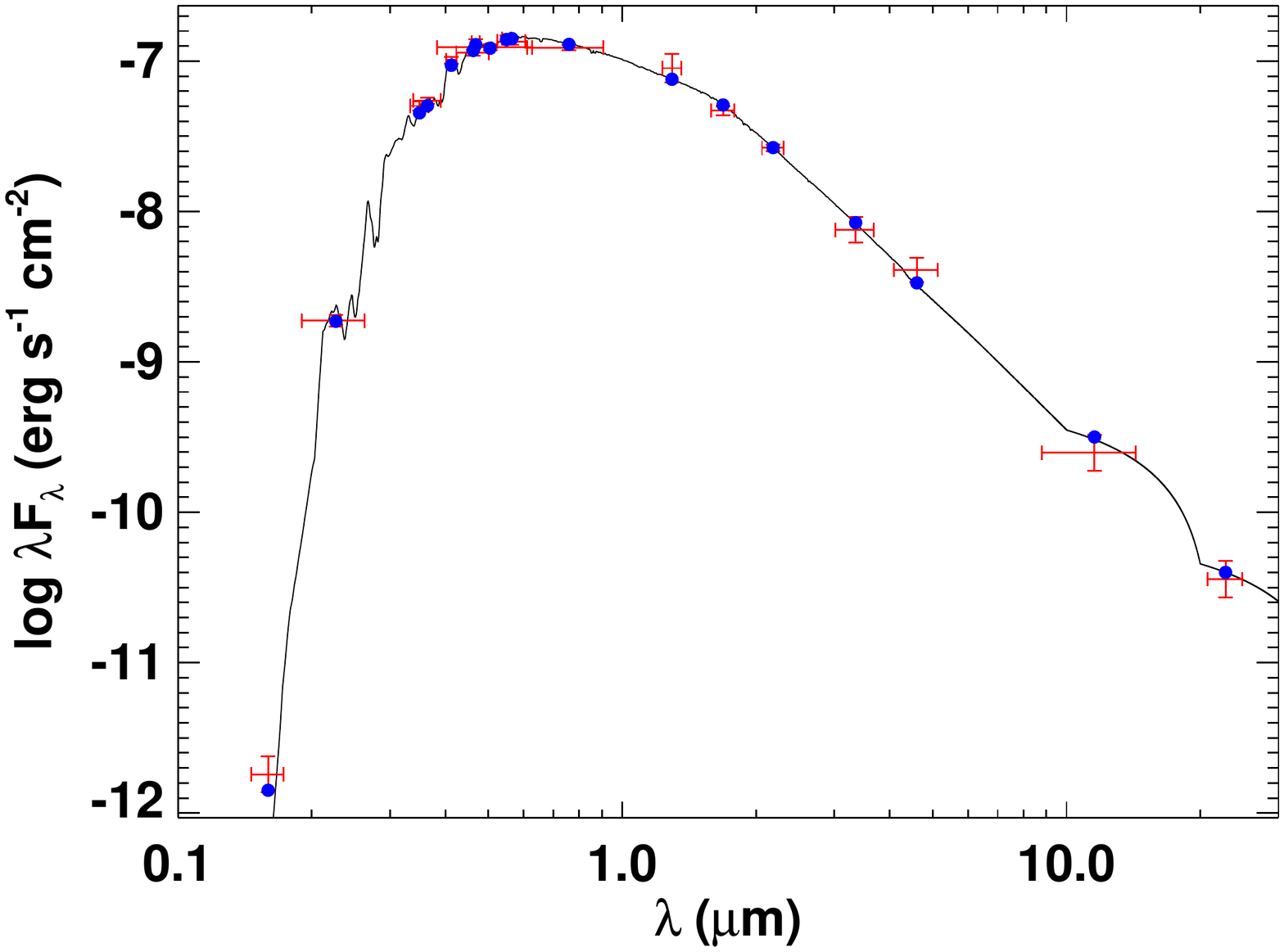} 
 \centering\includegraphics[width=\columnwidth,trim=150 70 80 70,clip]{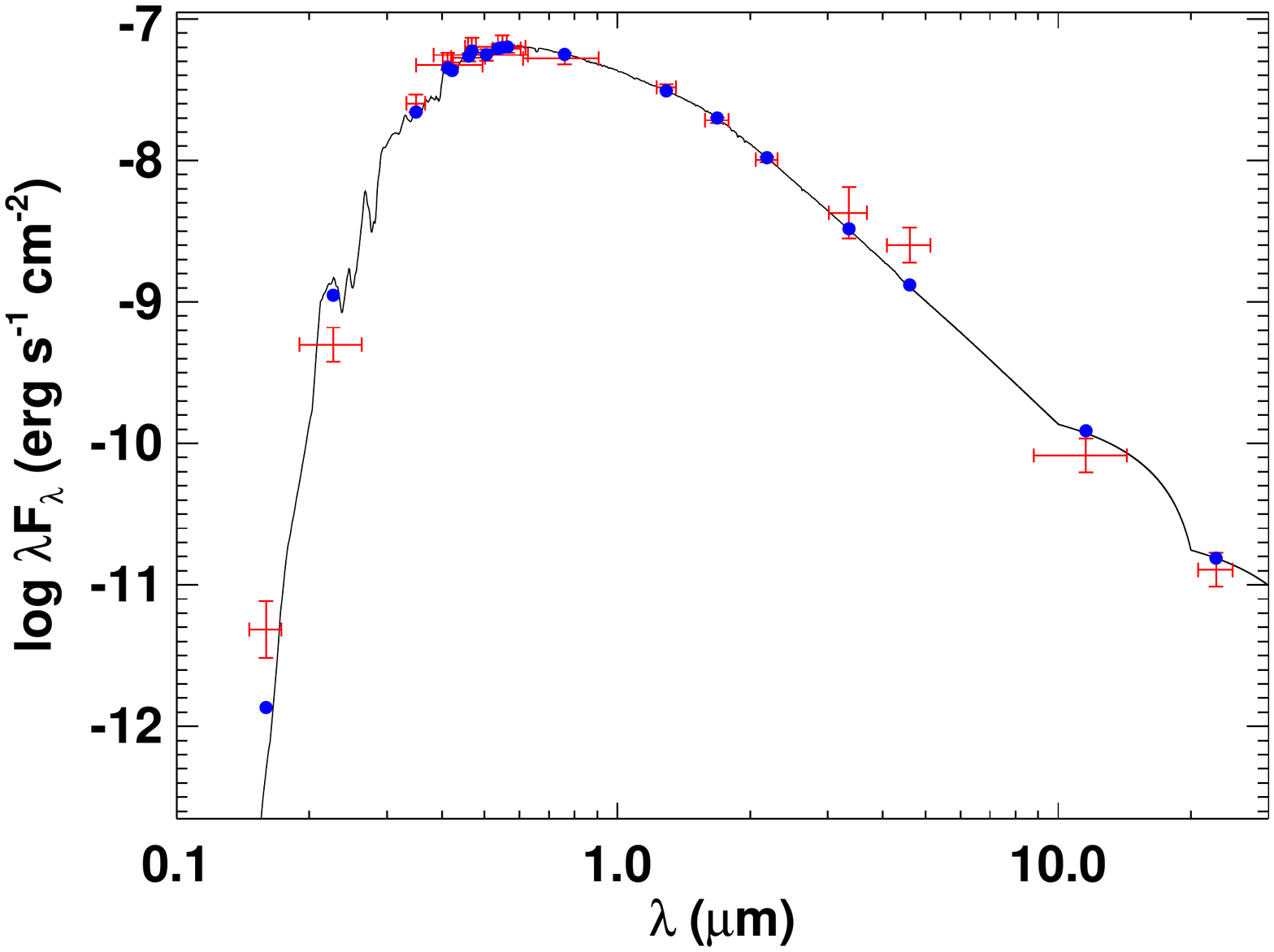} 
 \caption{Spectral energy distributions for $\rho$~CrB (top) and 88~Leo (bottom). Red symbols are the observed photometric measurements, where the horizontal bars represent the effective width of the passband. Blue symbols are the model fluxes from the best-fit Kurucz atmosphere model (black).\label{fig:sed}}
 \end{figure}

In order to provide an initial, empirical constraint on the stellar luminosities and radii, we performed an analysis of the broadband spectral energy distributions (SEDs) together with the {\it Gaia\/} EDR3 parallaxes following the procedures described in \citet{Stassun:2016} and \citet{Stassun:2017,Stassun:2018}. We pulled the FUV and NUV fluxes from {\it GALEX}, the $UBV$ magnitudes from \citet{Mermilliod:2006}, the Str\"omgren $uvby$ magnitudes from \citet{Paunzen:2015}, the $JHK_S$ magnitudes from {\it 2MASS}, the W1--W4 magnitudes from {\it WISE}, and the $G\,G_{BP}\,G_{RP}$ magnitudes from {\it Gaia}. Together, the available photometry spans the full stellar SED over the wavelength range 0.2--22~$\mu$m (see Figure~\ref{fig:sed}). 

We performed a fit using Kurucz stellar atmosphere models \citep{Castelli2004}, adopting the effective temperature ($T_{\rm eff}$) and metallicity ([M/H]) from the spectroscopically determined values of \citet{Brewer2016}. Uncertainties were inflated to account for a realistic systematic noise floor: $T_{\rm eff} = 5833 \pm 78$~K, $\text{[M/H]} = -0.18 \pm 0.07$~dex for $\rho$~CrB, and $T_{\rm eff} = 6002 \pm 78$~K, $\text{[M/H]} = +0.04 \pm 0.07$~dex for 88~Leo. The extinction ($A_V$) was fixed at zero due to the proximity of the stars to Earth. The resulting fits (Figure~\ref{fig:sed}) have a reduced $\chi^2$ between 1--2 for both stars. Integrating the (unreddened) model SED gives the bolometric flux at Earth ($F_{\rm bol}$). Taking this $F_{\rm bol}$ together with the {\it Gaia\/} EDR3 parallax, with no systematic adjustment \citep[e.g., see][]{StassunTorres:2021}, yields bolometric luminosities for $\rho$~CrB and 88~Leo of $L_{\rm bol} = 1.746 \pm 0.041\ L_\odot$ and $L_{\rm bol} = 1.482 \pm 0.088\ L_\odot$, respectively. In addition, the $L_{\rm bol}$ together with the $T_{\rm eff}$ yields stellar radii for $\rho$~CrB and 88~Leo of $R = 1.295 \pm 0.025\ R_\odot$ and $R = 1.127 \pm 0.037\ R_\odot$, respectively. Finally, we can estimate the stellar mass using the empirical eclipsing-binary based relations of \citet{Torres:2010}, which gives $M = 1.09 \pm 0.07\ M_\odot$ and $M = 1.14 \pm 0.07\ M_\odot$ for $\rho$~CrB and 88~Leo, respectively.

\subsection{X-ray data}\label{sec2.4} 

We obtained a {\it Chandra} observation of $\rho$~CrB  using the High Resolution Camera imaging detector (HRC-I) on 2020~Apr~19 starting at UT\,14:59 for a net exposure time of 11870\,s. This instrument was chosen because it has the best available low-energy sensitivity for imaging observations. An earlier observation of $\rho$~CrB had also been obtained (PI:~S.~Saar) several years earlier on 2012~Jan~17 beginning at UT\,13:12 using the Advanced CCD Imaging Spectrometer spectroscopic array (ACIS-S) on the back-illuminated CCD (``s3'') for a net exposure of 9835\,s.  

Both observations were downloaded from the {\it Chandra} archive and reprocessed using the {\it Chandra} Interactive Analysis of Observations (CIAO) software version 4.13 and calibration database version 4.9.4. While the ACIS-S data in principle have energy information for each photon from which a low-resolution X-ray spectrum can be derived, the $\rho$~CrB data contained only a handful of photon counts. The HRC-I data have no useful energy resolution. Analysis for both detectors therefore proceded similarly, by examining the photon counts attributable to $\rho$~CrB and using the instrument effective area to infer the implications for the X-ray flux. A summary of the observational results is presented in Table~\ref{tab1}.

\begin{table}[b]
\setlength{\tabcolsep}{4pt} 
\centering 
\caption{Summary of {\it Chandra} results for $\rho$~CrB}
\begin{tabular}{lcc}
\hline\hline
Parameter & HRC-I & ACIS-S\\
\hline 
Chandra ObsID & 22308 & 12396 \\
Net exposure (s) & 11870 & 9835 \\
$\rho$~CrB count rate  (count~ks$^{-1}$) & $2.85\pm 0.51$ & $0.77\pm 0.31$ \\
\cline{2-3}
Isothermal plasma temperature & 
\multicolumn{2}{l}{ $(1.58\pm 0.32) \times 10^6$~K} \\
X-ray Luminosity $L_X^a$ & \multicolumn{2}{l}{$(9.1\pm 1.9)\times 10^{26}$  erg s$^{-1}$}\\
\hline
\end{tabular} 
\label{tab1} \\
{\footnotesize $^a$Best estimate of the X-ray luminosity assuming an isothermal optically-thin plasma radiative loss model with a solar mixture of abundances scaled by a metallicity [M/H]\,$=-0.18$, and an interstellar absorbing column of $1.95\times 10^{18}$~cm$^{-2}$.\vspace*{-12pt}}
\end{table} 

In order to provide insight into the source X-ray luminosity giving rise to the HRC-I and ACIS-S signals, we used the {\tt PIMMS} software\footnote{\url{https://heasarc.gsfc.nasa.gov/docs/software/tools/pimms.html}} version 4.11 to convert the observed HRC-I and ACIS-S count rates to the incident \mbox{X-ray} flux. Since we are lacking counts in the ACIS-S data to estimate a coronal temperature, we made the flux conversion for a range of isothermal plasma temperatures. We adopted the APEC optically-thin plasma radiative loss model \citep{Foster2012}, the metallicity of [M/H]=$-$0.18 from \citet{Brewer2016}, and the solar abundance mixture of \citet{Asplund2009}, together with an intervening hydrogen column density of $1.95\times 10^{18}$~cm$^{-2}$. This column density was estimated by interpolation within the compilations of column density measurements of \citet{Gudennavar2012} and \citet{Linsky2019}, for the {\it Gaia} EDR3 distance of 17.51~pc.

\begin{figure}
 \centering\includegraphics[width=\columnwidth]{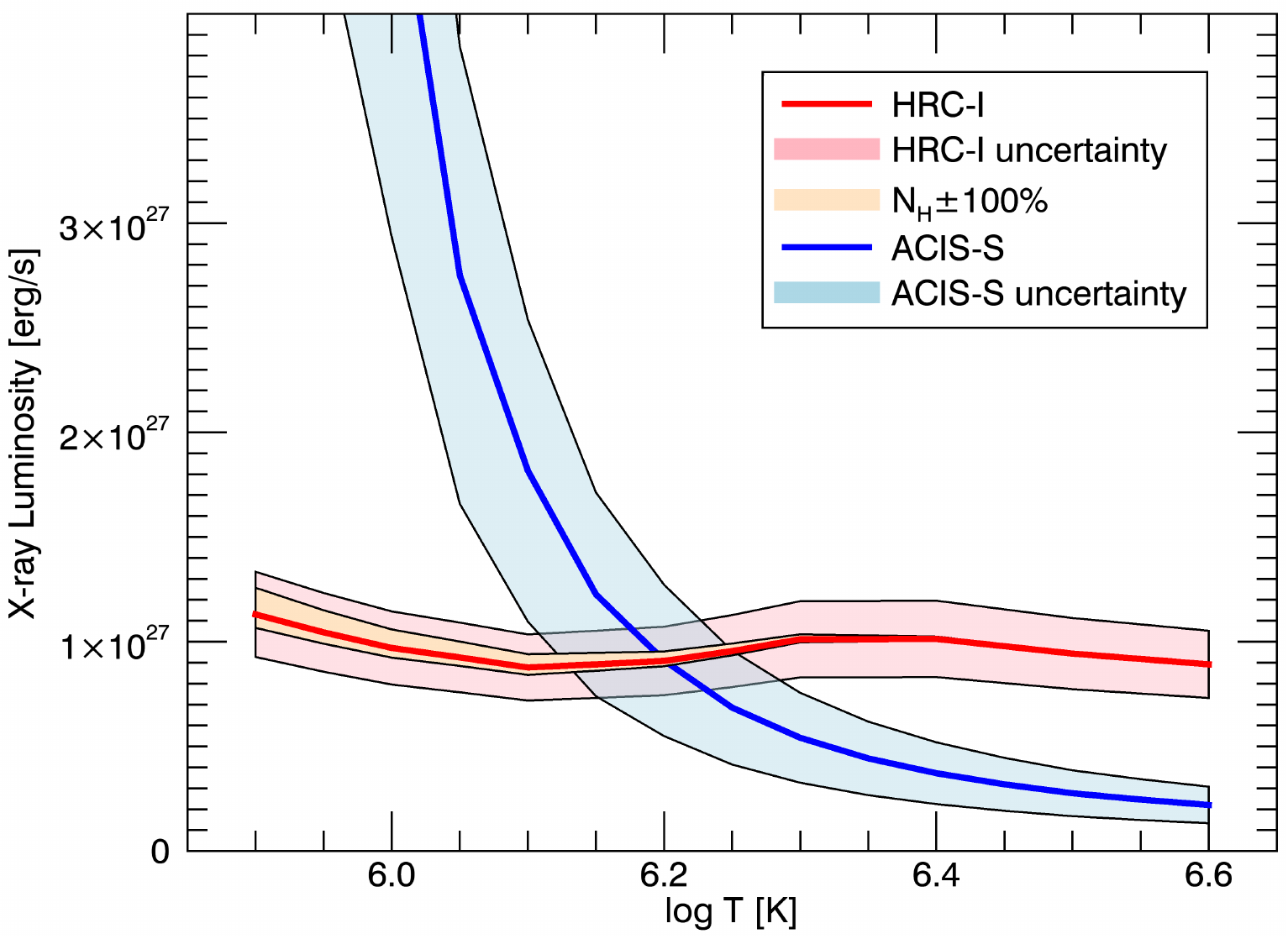} 
 \caption{The X-ray luminosity of $\rho$~CrB for isothermal, collision-dominated, optically-thin plasma radiative loss as a function of plasma temperature implied by the observed {\it Chandra} HRC-I  and ACIS-S count rates. Shaded regions indicate the uncertainties arising from the count rate measurements and a 100\% error in the assessment of the intervening neutral hydrogen column density.
 \label{fig3}}
 \end{figure}

The X-ray luminosities in the ROSAT 0.1--2.4\,keV band corresponding to the observed HRC-I and ACIS-S count rates are illustrated as a function of isothermal plasma temperature in Figure~\ref{fig3}. Shaded regions illustrate the range of uncertainties based on the uncertainties in the extracted count rates. Sensitivity of the results to the adopted absorbing column was determined by repeating the luminosity calculations for lower and higher values of $N_\mathrm{H}$ by a factor of two. Sensitivity to metallicity was also checked in a similar way, by varying metallicity by a factor of two, and found to be negligible.

Table~\ref{tab1} summarises the {\it Chandra} results for coronal luminosity and plasma temperature under the isothermal approximation. Final values were determined by the intersection of the HRC-I and ACIS-S $L_X$-$T$ loci and uncertainty ranges. By far the largest uncertainty is in the estimate of the X-ray count rates. The final estimate of the X-ray luminosity for $\rho$~CrB, $(9.1 \pm 1.9)\times 10^{26}$~erg~s$^{-1}$, is very similar to that of the quiet Sun \citep[e.g.][]{Judge2003}, while the temperature is similar to the peak of the quiet Sun emission measure distribution \citep[e.g.][]{Brosius96}.

For 88~Leo, we start with the X-ray luminosity $\log L_X = 27.77$ from \citet{Wright+2011}, which was derived from ROSAT PSPC data. This value was computed using the observed count rate \citep[$0.0306\pm0.0102$~counts~s$^{-1}$,][]{Voges+2000} and the hardness ratio (HR\,=\,$-1$) calibration of \citet{Schmitt+1995} with a distance of $d = 23.0$~pc and $L_{\rm bol} = 1.50\ L_{\odot}$. Adjusting for the updated properties from Section~\ref{sec2.3}, we find $L_X\,=\,(6.1 \pm 2.1)\times 10^{27}$~erg~s$^{-1}$ and $\log L_X/L_{\rm bol}\,=\,-5.96^{+0.14}_{-0.19}$. Adopting the SED radius from Section~\ref{sec2.3}, the surface flux is $\log F_X\,=\,4.90^{+0.14}_{-0.19}$. As a check, we used {\tt PIMMS} with the above count rate, a column of $N_H\,=\,10^{18}$~cm$^{-2}$ \citep[based on the star's presence in the NGP cloud,][]{Linsky2019}, solar abundance and APEC models. We find $\log F_X\,=\,4.92$ at $T_X\,=\,1.04^{+0.75}_{-0.35}\times 10^6$~K, which is reasonable given the hardness ratio. The ROSAT observations were obtained in late 1990, when the Ca~HK emission was slightly below the average level, so the \mbox{X-ray} observations should represent below average coronal conditions for 88~Leo.

\section{Asteroseismology of \texorpdfstring{$\rho$~C\lowercase{r}B}{rho CrB}}\label{sec3}

\subsection{Global oscillation parameters}\label{sec3.1} 

\begin{figure}
 \centering\includegraphics[width=\columnwidth]{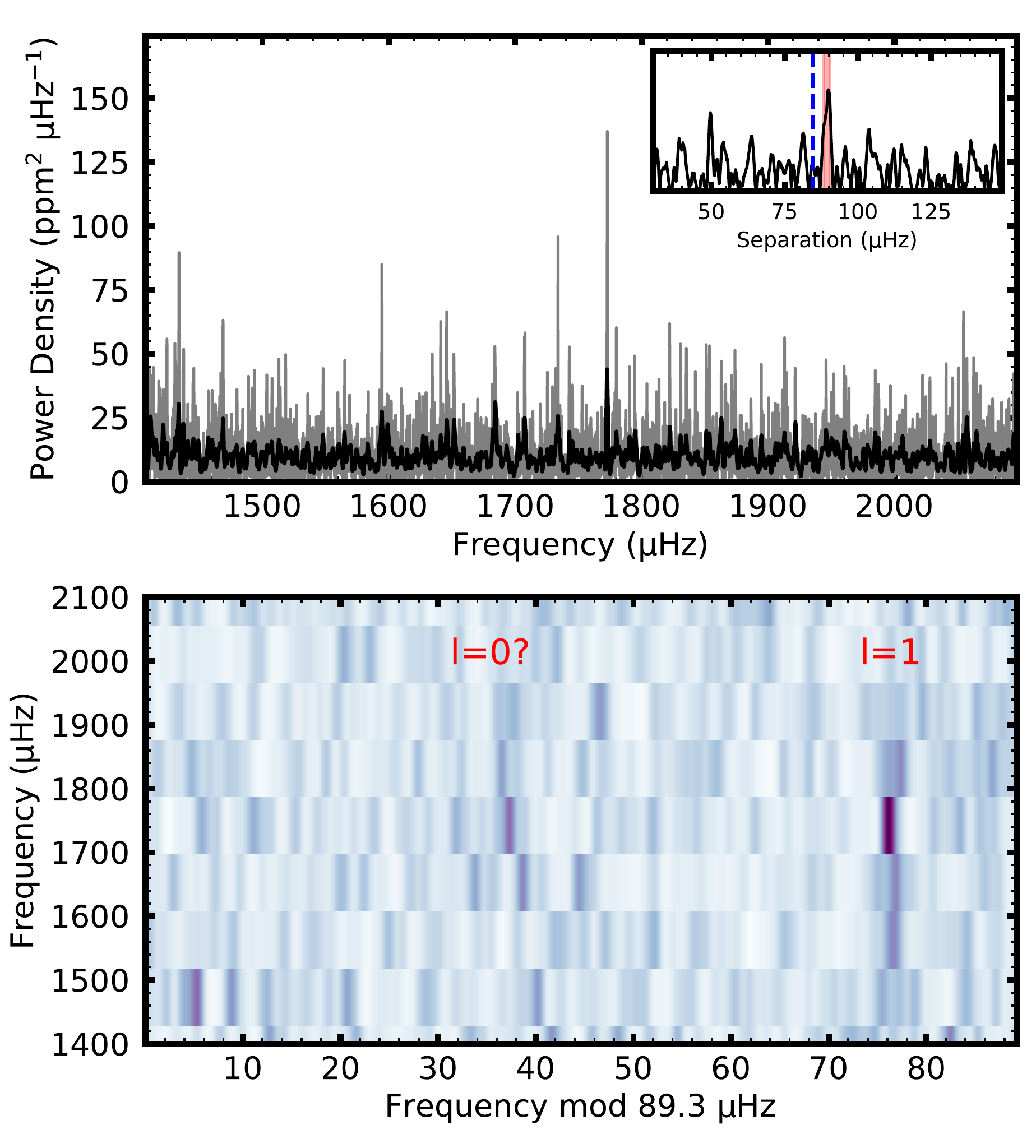} 
 \caption{Top panel: Power density spectrum of the TESS light curve for $\rho$~CrB. The inset shows an autocorrelation of the power spectrum, with the expected large frequency separation marked by a vertical dashed line. The red shaded area marks the measured large separation. Bottom panel: \'{E}chelle diagram of the power spectrum in the top panel.\label{fig4}}
 \end{figure}

The expected frequency of maximum power ($\nu_{\rm max}$) for $\rho$~CrB based on the TESS Asteroseismic Target List \citep[ATL,][]{schofield19} is $\approx$\,2000\,$\mu$Hz, with a detection probability of $\approx$\,65\%. The top panel of Figure~\ref{fig4} shows a power spectrum of the TESS light curve from Section~\ref{sec2.2}. The spectrum displays a low signal-to-noise power excess around 1800\,$\mu$Hz, which is consistent with the ATL given uncertainties in predicted $\nu_{\rm max}$ values.

To test whether the power excess is consistent with solar-like oscillations, we calculated an autocorrelation of the power spectrum between $\approx$\,1400-2100\,$\mu$Hz (inset in the top panel of Figure~\ref{fig4}). The autocorrelation shows a peak at $\approx$\,89\,$\mu$Hz, close to the expected value for the characteristic large frequency separation ($\Delta\nu$) for solar-like oscillations in this frequency range \citep{stello09c}. We furthermore calculated an \'{e}chelle diagram (bottom panel of Figure~\ref{fig4}) by dividing the power spectrum into equal segments with length $\Delta\nu$ and stacking one above the other, so that modes with a given spherical degree align vertically in ridges \citep{grec83}. The offset of the visible ridge in the \'{e}chelle diagram, which is sensitive to the properties of the near-surface layers of the star \citep[e.g.][]{jcd14}, is consistent with expectations for a ridge of dipole ($l=1$) modes based on \textit{Kepler} measurements of stars with similar $\Delta\nu$ and $T_{\rm eff}$ \citep{white11}.

We used several independent methods \citep{huber09,mosser09,mathur10b,campante18} to extract global oscillation parameters from the power spectrum, which yielded broadly consistent results. Estimates of $\nu_{\rm max}$ showed a large spread, as expected for a low signal-to-noise detection \citep{chaplin14}, so we did not adopt a constraint for our subsequent analysis. We adopted $\Delta\nu = 89.3 \pm 1.1\ \mu$Hz as measured by the SYD pipeline, which was consistent with measurements from other methods.

We also searched for oscillations in 88~Leo, which yielded a null detection. The star is not included in the ATL, but based on the stellar properties from Section~\ref{sec2} it is less evolved than $\rho$~CrB with an expected $\nu_{\rm max}$ of $\approx 2700$\,$\mu$Hz and a detection probability $\approx$\,26\%. Given the low S/N detection in $\rho$~CrB, the fainter apparent magnitude of 88~Leo, and the fact that oscillation amplitudes decrease with increasing $\nu_{\rm max}$ and with higher activity \citep{Garcia2010, Chaplin2011, Mathur2019}, we conclude that the null detection is consistent with expectations.

\subsection{Grid-based modeling} 

Grid-based modeling of $\rho$~CrB was performed using the Yale-Birmingham pipeline \citep{Basu2010, Basu2012, Gai2011} with $\Delta\nu$, [M/H], $T_{\rm eff}$, and luminosity as inputs, and the results are listed in Table~\ref{tab2}. The search was conducted on a grid containing two sub-grids---one with the solar-calibrated mixing length parameter, and the second using a metallicity-dependent mixing length, with the dependence given by \citet{Viani2018}. Both sub-grids assume a linear relation $\Delta Y/\Delta Z \approx 1.5$ that was obtained using a calibrated solar model assuming a primordial helium abundance of $0.248$ \citep{Steigman2010}. The grids have models with masses between 0.7\,$M_\odot$ and 3.3\,$M_\odot$ in intervals of 0.025\,$M_\odot$, evolved from the zero-age main-sequence to nearly the tip of the red-giant branch. The models were constructed with metallicities ranging from [M/H]=$-2.4$ to $+0.5$. The metallicity grid has a spacing of 0.1~dex between $-2.0$ and $+0.5$, and a spacing of 0.2~dex at lower metallicity. The metallicity scale is that of \citet{Grevesse1998}, i.e., $\text{[M/H]}=0$ corresponds to $Z/X=0.023$.

The grids were constructed using the Yale Stellar Evolution Code \citep[YREC,][]{Demarque2008} for consistency with the rotational evolution modeling in Section~\ref{sec5}. The models were constructed using OPAL opacities \citep{Iglesias1996} supplemented with low temperature opacities from \citet{Ferguson2005}. The OPAL equation of state \citep{Rogers2002} was used. All nuclear reaction rates were obtained from \citet{Adelberger1998}, except for that of the $^{14}N(p,\gamma)^{15}O$ reaction, for which we adopted the rate of \citet{Formicola2004}. All models included gravitational settling of helium and heavy elements using the formulation of \citet{Thoul1994}, with the diffusion coefficients smoothly decreased for stars more massive than $1.25\ M_\odot$. The large separation $\Delta\nu$ for the models were calculated from the frequencies of their radial modes, which in turn were calculated with the code of \citet{Antia1994}. The large separations were corrected for the surface term by applying the correction obtained by \citet{Viani2019}. 

 \begin{deluxetable*}{lccccc}
 \tablecaption{Stellar Properties of $\rho$~CrB and 88~Leo\label{tab2}}
 \tablehead{ & \multicolumn{2}{c}{$\rho$~CrB} & & & \\ 
 \cline{2-3}
 \colhead{} & \colhead{Asteroseismic} & \colhead{Other} & \colhead{~~~~~} & \colhead{88~Leo} & \colhead{Source}}
 \startdata
 $\log R'_{\rm HK}[T_{\rm eff}]$ (dex) & $\cdots$    & $-5.177 \pm 0.015$ & & $-4.958 \pm 0.015$ & (1) \\
 $P_{\rm rot}$ (days)   & $\cdots$                   & $20.3 \pm 1.8$     & & $15.0 \pm 0.3$     & (1) \\
 $T_{\rm eff}$ (K)      & $5817^{+32}_{-33}$         & $5833 \pm 78$      & & $6002 \pm 78$      & (2) \\
 $[$M/H$]$ (dex)        & $-0.19 \pm 0.06$           & $-0.18 \pm 0.07$   & & $+0.04 \pm 0.07$   & (2) \\
 $\log g$ (dex)         & $4.190 \pm 0.008$          & $4.29 \pm 0.08$    & & $4.38 \pm 0.08$    & (2) \\
 Radius ($R_\odot$)     & $1.304 \pm 0.012$          & $1.295 \pm 0.025$  & & $1.127 \pm 0.037$  & (3) \\
 Luminosity ($L_\odot$) & $1.749^{+0.036}_{-0.040}$  & $1.746 \pm 0.041$  & & $1.482 \pm 0.088$  & (3) \\
 Mass ($M_\odot$)       & $0.96 \pm 0.02$            & $1.09 \pm 0.07$    & & $1.14 \pm 0.07$    & (3) \\
 Age (Gyr)              & $9.8^{+0.7}_{-0.5}$        & $3.5 \pm 0.6$      & & $2.4 \pm 0.4$      & (4) \\
 $L_X$ ($10^{27}$~erg~s$^{-1}$) & $\cdots$           & $0.91 \pm 0.19$    & & $6.1 \pm 2.1$      & (5) \\
 \enddata
 \tablerefs{(1) \S\,\ref{sec2.1}; (2) \cite{Brewer2016}; (3) \S\,\ref{sec2.3}; (4) \cite{Barnes2007}; (5) \S\,\ref{sec2.4}}
 \end{deluxetable*}

\vspace*{-12pt}
\section{Magnetic Evolution}\label{sec4} 

\subsection{Activity-age relation}

 \begin{figure}
 \centering\includegraphics[width=\columnwidth,trim=0 0 30 40,clip]{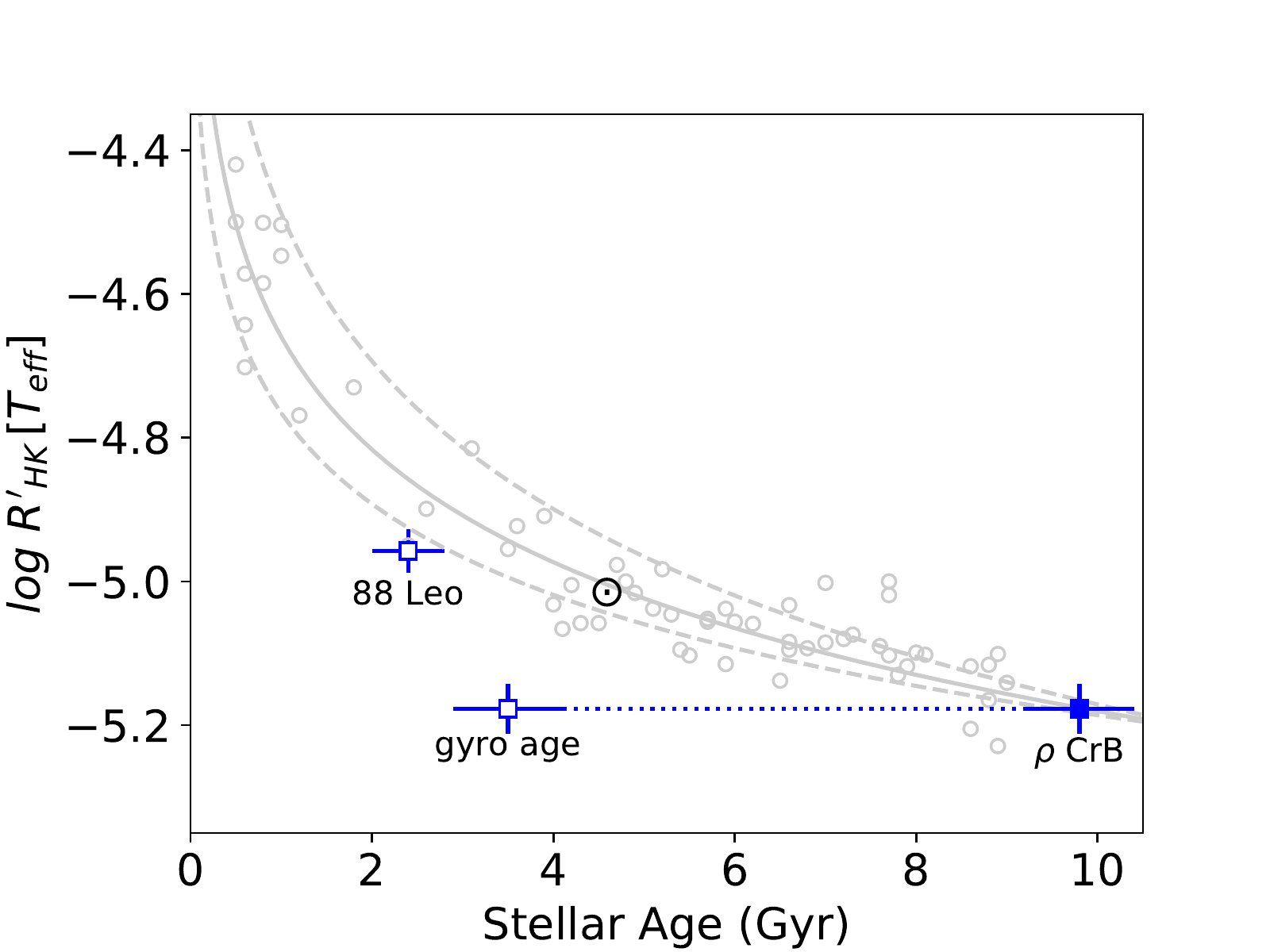} 
 \caption{Chromospheric activity versus stellar age for a sample of spectroscopic 
 solar twins from \cite{LorenzoOliveira2018} with ages determined from isochrone 
 fitting (gray circles). The asteroseismic age of $\rho$~CrB from TESS is 
 overplotted as a solid square, while updated ages from gyrochronology for both 
 stars are shown with open squares.\label{fig5}}
 \end{figure} 

Considering the stellar properties determined above, we can now evaluate how the age expected from the chromospheric activity level compares to other age indicators. To facilitate this comparison, in Figure~\ref{fig5} we show the activity-age relation for a sample of spectroscopic solar twins from \cite{LorenzoOliveira2018}. The ages for this sample (gray circles) were determined from isochrone fitting, and the chromospheric activity scale was calibrated using $T_{\rm eff}$ rather than B$-$V color. The derived activity-age relation with uncertainties (gray lines) should be applicable to stars that have a mass and metallicity similar to the Sun. We can place other stars on this same activity scale using their spectroscopic $T_{\rm eff}$ and average $S$-index, with a small correction for non-solar metallicity \citep[{0.213$\times$[M/H]},][]{SaarTesta2012}. The horizontal error bars indicate the age uncertainty, while the vertical error bars reflect the uncertainties in $T_{\rm eff}$ and [M/H].

Using these procedures to place $\rho$~CrB and 88~Leo on the chromospheric activity scale for solar twins, we can evaluate their ages from asteroseismology and gyrochronology. The asteroseismic age for $\rho$~CrB from Table~\ref{tab2} is shown as a solid square in Figure~\ref{fig5}, which falls directly on the activity-age relation for solar twins. Although we were unable to determine an asteroseismic age for 88~Leo, the age from gyrochronology should be reliable for this star because it is not yet below the critical activity level where weakened magnetic braking is inferred \citep{vanSaders2016, Brandenburg2017}. The updated gyrochronology ages for both stars are indicated with open squares \citep{Barnes2007}, showing a reasonable agreement for 88~Leo considering its higher mass \citep[$M_{\rm iso}=1.10\ M_\odot$,][]{Valenti2005} but revealing a strong disagreement for $\rho$~CrB. In Section~\ref{sec5}, we examine this tension in greater detail.

\subsection{Magnetic Braking Torque}

We can estimate the strength of magnetic braking for $\rho$~CrB and 88~Leo by combining the wind modeling prescription of \cite{Finley2017, Finley2018} with the constraints on magnetic morphology from \cite{Metcalfe2019}. Given the polar strengths of an axisymmetric dipole, quadrupole, and/or octupole magnetic field, along with the mass-loss rate, rotation period, stellar mass and radius, this prescription yields an estimate of the magnetic braking torque based on analytical fits to a set of detailed magnetohydrodynamic wind simulations. Although 88~Leo exhibits a nonaxisymmetric polarization profile, the amplitude of the signal can be reproduced with an axisymmetric dipole having a polar field strength $B_{\rm d}=-5$~G. For $\rho$~CrB, \cite{Metcalfe2019} cite upper limits on the polar field strength assuming a pure axisymmetric dipole ($|B_{\rm d}|\le0.7$~G) or quadrupole field ($|B_{\rm q}|\le2.4$~G), with the latter being larger due to geometric cancellation effects. An identical analysis of the same LBT data yields an upper limit on a pure axisymmetric octupole field of $|B_{\rm o}|\le19.6$~G. However, the LBT observations also showed that the disk-integrated line-of-sight magnetic field in $\rho$~CrB is about 64\% as strong as in 88~Leo, which agrees well with the relative chromospheric activity levels listed in Table~\ref{tab2}. Given the upper limit on the dipole component, the global field of $\rho$~CrB appears to be dominated by quadrupolar and higher-order components to account for its relative line-of-sight field and activity level.

Observationally, the mass-loss rate is one of the least certain quantities required by the wind modeling prescription. If we initially fix the mass-loss rate to the solar value for both stars ($\dot{M}_\odot=2\times10^{-14}\ M_\odot/\text{yr}$) and adopt the stellar properties from Table~\ref{tab2}, we find that the magnetic braking torque for $\rho$~CrB is $\la$20\% as strong as for 88~Leo. This estimate does not depend strongly on whether we adopt the asteroseismic or other estimates of radius and mass for $\rho$~CrB, so we adopt the asteroseismic properties for further analysis. The mass-loss rate generally decreases with stellar age, so we might expect it to be larger than the solar value at the updated gyrochronology age of 88~Leo (2.4~Gyr), and smaller by the asteroseismic age of $\rho$~CrB (9.8~Gyr). 

If we adopt the scaling relation $\dot{M}\propto F_X^{0.77}$ from \cite{Wood2021} and calculate the X-ray fluxes from the luminosities in Section~\ref{sec2.4}, the mass-loss rate changes from 2.0\,$\dot{M}_\odot$ to 0.36\,$\dot{M}_\odot$ between the ages of these two stars\footnote{If we adopt the steeper scaling relation $\dot{M}\propto F_X^{1.29}$ of \cite{Wood2018} derived from GK dwarfs only, the mass-loss rate estimates become 3.1\,$\dot{M}_\odot$ for 88~Leo and 0.18\,$\dot{M}_\odot$ for $\rho$~CrB.}, and the magnetic braking torque for $\rho$~CrB becomes $\la$8\% as strong as for 88~Leo. We can estimate the relative contributions to this total reduction in magnetic braking torque by changing the parameters of the 88~Leo wind model one at a time to the values in the $\rho$~CrB model.
The largest factor that contributes to the reduction in magnetic braking torque is the shift in morphology towards quadrupolar and higher-order fields ($-$67\% from shifting the field from pure dipole to pure quadrupole), followed by the evolutionary change in mass-loss rate ($-$60\%), with smaller contributions from the weaker magnetic field (up to $-$34\% from changing the strength of a quadrupole field from 5~G to 2.4~G) and slower rotation ($-$26\%). The slightly lower mass (+4\%) and evolutionary change in the radius (+58\%) actually increase the relative magnetic braking torque, masking some of the other effects.

\vspace*{12pt}
\section{Rotational Evolution}\label{sec5} 

We modeled the rotational evolution of $\rho$~CrB using the methodology laid out in \citet{Metcalfe2020}. We assumed solid body rotation, and used the \texttt{rotevol} \citep{Somers2017} tracer code to track the angular momentum evolution as a function of time, given a set of YREC evolutionary tracks and interpolation tools in \texttt{kiauhoku} \citep{Claytor2020}. We used the same model grid as that in \citet{Metcalfe2020} and adopted the same braking law parameters, with minor changes that we describe here. We scaled the critical Rossby number, Ro$_{\rm crit}$ in terms of the solar value, since the \citet{vanSaders2016} model grid and our current grid have slightly different solar calibrations due to differing input physics. We adopted Ro$_{\rm crit} = 0.92~\textrm{Ro}_{\odot}$ as estimated in \citet{vanSaders2019}. Second, although unimportant for the late time rotational evolution, we chose a constant specific angular momentum (cm$^2$~s$^{-1}$) of $\log{j_{spec}} = 16.3$ dex at 10~Myr \citep{Somers2017} as our initial condition.

We utilized the same Monte Carlo approach as in \citet{Metcalfe2020} in which the mass, initial metallicity, age, and mixing length are parameters of the model, with the asteroseismic radius and the spectroscopic surface [M/H] and $T_{\rm eff}$ as the observables. We adopted strict Gaussian priors on the mass ($0.96\pm0.02~M_{\odot}$) and age ($9.8\pm0.8$ Gyr) from the asteroseismic analysis, and a broader prior on the mixing length ($1.8\pm0.3$). In both cases, the rotation period is a prediction of the model, rather than a parameter we use in the fit itself. We used 8 walkers, each running for 100,000 steps.

\begin{figure}
 \centering\includegraphics[width=\columnwidth,trim=7 7 0 0,clip]{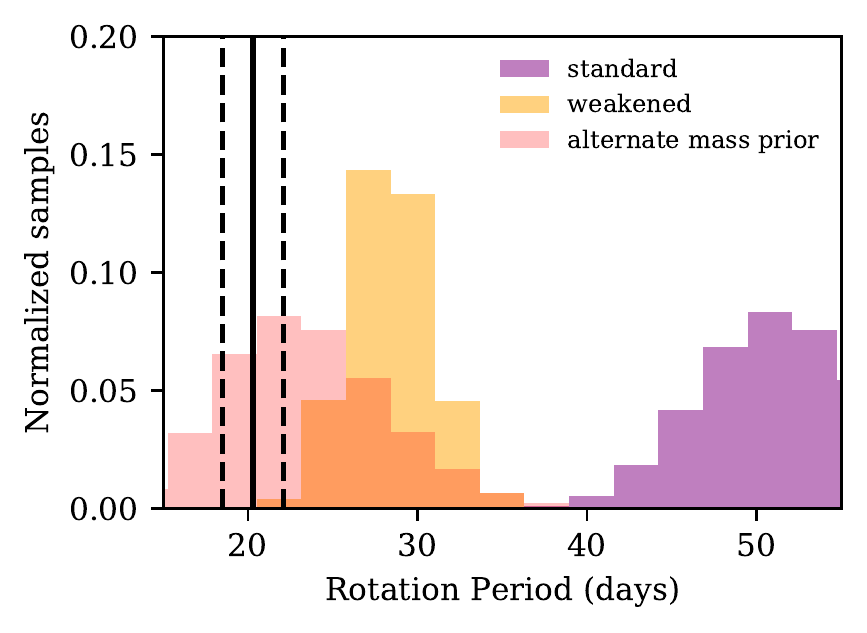} 
 \caption{Predictions from a standard spin-down model (purple), weakened braking model (orange), and weakened braking model with a mass prior of $1.09\pm0.07~M_{\odot}$ (pink) for the rotation period of $\rho$~CrB. The observed rotation period from Section~\ref{sec2.1} is shown with black vertical lines.\label{fig6}}
 \end{figure}

The standard spin-down model predicts a rotation period of $52 \pm 5$ days for $\rho$~CrB, while the weakened braking model with Ro$_{\rm crit} = 0.92~\textrm{Ro}_{\odot}$ predicts a rotation period of $28 \pm 2$ days. We show in Figure~\ref{fig6} the posteriors on the predicted rotation distributions for both the standard spin-down and weakened magnetic braking cases in comparison to the observed period: both models predict longer periods. 

We verified that changing the initial angular momentum is insufficient to relieve the tension, as in \citet{Metcalfe2020}. Similarly, allowing the model to deviate from purely solid body rotation is also unlikely to result in more rapid rotation: in both the Sun and asteroseismic samples the rotation with depth is consistent with a solid body \citep{Deheuvels2020}. The convection zone of $\rho$~CrB has not yet begun to deepen at its current position in the HR diagram, and it is unlikely to be dredging up higher angular momentum material from a differentially rotating interior, even if such radial shear exists. Furthermore, when the core and envelope are allowed to decouple rotationally \citep{MacGregor1991} the surface rotation rate tends to be \textit{slower} than a solid body model, because wind-driven loss drains angular momentum from the smaller, decoupled reservoir of the convective envelope. This star is also still hot enough ($\sim$5800~K) that assumptions about the convective mixing length have a comparatively mild effect on the predicted period.  

An underestimated stellar mass would result in predicted rotation periods that are too long, and indeed there is moderate tension between the asteroseismic mass and the empirical mass scale from eclipsing binaries. If we instead adopt a mass prior of $1.09 \pm 0.07~M_\odot$ (while also adopting an uninformative age prior) we predict a period of $23^{+5}_{-4}$ days for the weakened magnetic braking case. The inferred mass is $1.00\pm0.03\ M_\odot$ \citep[consistent with the isochrone mass,][]{Valenti2005,Brewer2016}, with predicted properties within 1$\sigma$ of the observed $L$, $R$, $T_{\rm eff}$, and surface [M/H]. The increased mass does require an age younger by about 2~Gyr (also consistent with isochrone estimates), but this is unsurprising: on the subgiant branch (SGB) near the turnoff, the age is tightly correlated with model mass. The ages of such stars are essentially equal to the main-sequence lifetime, and their rotational evolution shifts from being strongly dependent on time to strongly dependent on the structural evolution across the SGB.

We applied the same modeling techniques to 88 Leo, and find excellent agreement with the observed rotation period in both standard and weakened braking prescriptions. The predicted period is $15\pm2$ days for both models: they do not differ significantly because the Rossby number of 88 Leo is approximately equal to our adopted critical Rossby number ($\textrm{Ro} = 1.0\pm0.1\ \textrm{Ro}_{\rm crit}$). Both the standard and weakened braking models are identical until the critical Rossby number is exceeded, and thus both predict the same rotation period for 88 Leo.

\section{Summary and Discussion}\label{sec6} 

By combining archival stellar activity data from MWO with asteroseismology from TESS, we have probed the nature of the transition that appears to decouple the evolution of rotation and magnetism in middle-aged stars. We characterized two stars ($\rho$~CrB and 88~Leo) with activity levels on opposite sides of the proposed mid-life transition---verifying their mean activity levels and rotation periods (Section~\ref{sec2.1}), quantifying their X-ray luminosities to estimate mass-loss rates (Section~\ref{sec2.4}), and deriving precise asteroseismic properties for the post-transition star $\rho$~CrB (Section~\ref{sec3}). Analysis of the resulting observational constraints reveals that the asteroseismic age of $\rho$~CrB agrees with the expected evolution of its mean activity level, while the age from gyrochronology does not (Figure~\ref{fig5}). No such tension exists for 88~Leo, suggesting a divergence in the evolution of rotation and magnetism between 2.4 and 3.5~Gyr for stars with shallower convection zones than the Sun. 

Using a simple wind modeling prescription with previously published spectropolarimetric constraints on the global magnetic fields \citep{Metcalfe2019}, we find that the magnetic braking torque for $\rho$~CrB is more than an order of magnitude smaller than for 88~Leo, primarily due to a shift in morphology toward smaller spatial scales but reinforced by the evolutionary change in mass-loss rate and other properties (Section~\ref{sec4}). Rotational evolution models adopting standard spin-down can match the observational constraints for 88~Leo, but they fail for $\rho$~CrB. By contrast, models with weakened magnetic braking can more readily explain the fast rotation of $\rho$~CrB, particularly if the asteroseismic properties are slightly biased from the relatively low S/N detection (Figure~\ref{fig6}).

Future TESS observations may allow refinement of the stellar properties for $\rho$~CrB, and could yield an asteroseismic detection for 88~Leo. Both targets will be observed with 20-second cadence during Cycle~4, which yields a 20\% longer effective integration time due to the absence of onboard cosmic ray rejection, and also avoids significant attenuation of signals near the Nyquist frequency of 2-minute sampling \citep{Huber2021}. The latter is particularly important for 88~Leo ($\nu_{\rm max} \approx 2700~\mu$Hz), which will be observed during Sectors~45-46 (2021 Nov/Dec) and Sector~49 (2022 Mar), further improving the detection probability. Although 88~Leo has a K-dwarf companion separated by 15$\farcs$5, it only dilutes the signal from the primary by $\sim$10\%, and any solar-like oscillations in the K-dwarf are expected at a higher frequency and much lower amplitude. Additional observations of $\rho$~CrB will be obtained during Sector~51 (2022 May), and they can be combined with the Cycle~2 data to improve the S/N of the detection, potentially yielding a more precise value of $\Delta\nu$, a secure determination of $\nu_{\rm max}$, and perhaps some individual oscillation frequencies for detailed modeling. This may allow us to resolve the tension between the asteroseismic properties derived in Section~\ref{sec3} and the eclipsing binary mass scale, and possibly probe the impact of the observed non-solar abundance mixture for this star \citep{Brewer2016}.

Additional spectropolarimetic observations will provide new opportunities to test the mid-life transition hypothesis across a range of spectral types. Data recently obtained from the LBT include Stokes~V measurements of 18~Sco, 16~Cyg~A~\&~B, $\lambda$~Ser, and HD\,126053. The latter appears to be a transitional star like $\alpha$~Cen~A \citep{Metcalfe2017}, but with a rotation period and activity cycle very similar to the Sun. Such targets may offer the best constraints on the timescale for a shift in magnetic morphology, which must play out relatively quickly to explain the sudden reduction in magnetic braking torque suggested by observations \citep{vanSaders2016}. By contrast, evolutionary changes in the mass-loss rate, mean activity level, and rotation period (as a star expands on the main-sequence) should take place more gradually. Aside from 18~Sco \citep[which has a ground-based asteroseismic detection,][]{Bazot2011}, all of these targets will be observed by TESS with 20-second cadence in Cycle~4, and most of them have well-defined X-ray fluxes to constrain the mass-loss rates. Consequently, we should be able to extend the methodology applied above to a well-characterized sample of solar-type stars in the near future.

\vspace*{24pt}
The authors would like to thank Steven Cranmer, B.~J.\ Fulton, Sean Matt, Marc Pinsonneault, and Kaspar von~Braun for helpful exchanges.
T.S.M.\ acknowledges support from NSF grant AST-1812634, NASA grant 80NSSC20K0458, and Chandra award GO0-21005X. Computational time at the Texas Advanced Computing Center was provided through XSEDE allocation TG-AST090107. 
J.v.S.\ acknowledges support from  NASA grant 80NSSC21K0246. 
D.B.\ acknowledges support from NASA through the Living With A Star Program (NNX16AB76G) and from the TESS GI Program under awards 80NSSC18K1585 and 80NSSC19K0385. 
J.J.D.\ was supported by NASA contract NAS8-03060 to the {\it Chandra X-ray Center} and thanks the Director, Pat Slane, for continuing advice and support.
R.E.\ acknowledges NCAR for their support. The National Center for Atmospheric Research is sponsored by the National Science Foundation. 
D.H.\ acknowledges support from the Alfred P. Sloan Foundation, NASA grant 80NSSC21K0652, and NSF grant AST-1717000.
S.H.S.\ is grateful for support from NASA Heliophysics LWS grant NNX16AB79G, and HST grant HST-GO-15991.002-A.
W.H.B.\ acknowledges support from the UK Space Agency. 
T.L.C.\ is supported by Funda\c c\~ao para a Ci\^encia e a Tecnologia (FCT) in the form of a work contract (CEECIND/00476/2018).
A.J.F.\ is supported by the ERC Synergy grant ``Whole Sun'', \#810218.
O.K.\ acknowledges support by the Swedish Research Council, the Royal Swedish Academy of Sciences, and the Swedish National Space Agency.
S.M.\ acknowledges support from the Spanish Ministry of Science and Innovation with the Ramon y Cajal fellowship number RYC-2015-17697 and the grant number PID2019-107187GB-I00. 
T.R.\ acknowledges support from the European Research Council (ERC) under the European Union's Horizon 2020 research and innovation program (grant agreement No.\ 715947). 
V.S.\ acknowledges funding from the European Research Council (ERC) under the European Union's Horizon 2020 research and innovation program (grant agreement No.\ 682393 AWESoMeStars).
This work benefited from discussions within the international team ``The Solar and Stellar Wind Connection: Heating processes and angular momentum loss'' at the International Space Science Institute (ISSI).

\software{CIAO \citep[v4.13;][]{Fruscione2006}, PIMMS \citep[v4.11;][]{Mukai1993}, SYD \citep{huber09}, YREC \citep{Demarque2008}, rotevol \citep{Somers2017}, kiauhoku \citep{Claytor2020}}

\bibliographystyle{aasjournal}
\bibliography{references}

\end{document}